\documentclass[twocolumn,rmp]{revtex4}

\usepackage{dcolumn,graphicx,amsmath,amssymb}
\usepackage{txfonts}

\begin{document}

\title{The contact network of patients in a regional healthcare system}
\author{Fredrik Liljeros}
\affiliation{Department of Epidemiology and Biostatistics, Karolinska
  Institute, SE-171 77 Stockholm, Sweden}
\affiliation{Department of Sociology, Stockholm University 106 91
  Stockholm, Sweden}
\author{Petter Holme}
\affiliation{Department of Physics, University of Michigan, Ann Arbor,
  MI 48109, U.S.A.}
\author{Johan Giesecke}
\affiliation{Department of Epidemiology and Biostatistics, Karolinska
  Institute, SE-171 77 Stockholm, Sweden}

\begin{abstract}
Yet in spite of advances in hospital treatment, hospitals continue to
be a breeding ground for several airborne diseases and for diseases
that are transmitted through close contacts like SARS,
methicillin-resistant Staphylococcus aureus (MRSA), norovirus
infections and tuberculosis (TB). Here we extract contact networks for
up to 295,108 inpatients for durations up to two years from a database
used for administrating a local public healthcare system serving a
population of 1.9 million individuals. Structural and dynamical
properties of the network of importance for the transmission of
contagious diseases are then analyzed by methods from network
epidemiology. The contact networks are found to be very much
determined by an extreme (age independent) variation in duration of
hospital stays and the hospital structure. We find that that the
structure of contacts between in-patients exhibit structural
properties, such as a high level of transitivity~\cite{mejn:clu},
assortativity~\cite{mejn:ass} and variation in number of
contacts~\cite{and:may}, that are likely to be of importance for the
transmission of less contagious diseases.  If these properties are
considered when designing prevention programs the risk for and the
effect of epidemic outbreaks may be decreased.
\end{abstract}

\maketitle

\section{Limitations of traditional epidemiology}

A central parameter within infection epidemiology is the basic
reproduction number $R_0$~\cite{and:may}. $R_0$ is used to estimate
whether a disease is contagious enough to generate an epidemic in a
specific population. In its simplest form,  is defined as the expected
number of individuals that an infected individual will infect in a
completely susceptible population. If all individuals in a population
have approximately the same number of contacts, and the probability
that any pair of individuals will meet is equal, $R_0$  can be
estimated by the function below:
\begin{equation}\label{eq:r0}
  R_0=c\beta D ,
\end{equation}
where $c$ stands for umber of contacts per time unit, $\beta$ or
likelihood of passing on an infection per contact, and $D$ for the
average time an individual is infectious (measured in same time unit
as $c$).  To make an epidemic possible, the infected person must
infect more than one person on average. The threshold value for
epidemics is therefore $R_0=1$.

Studies have shown that $R_0$ in its simplest form may yield
misleading results. Anderson and May have demonstrated that a great
variation in number of contacts may compensate for a low average
number of contacts~\cite{and:may}. This is because individuals with
many contacts have a far greater probability of becoming infected, and
of passing on an infection. Therefore, in populations with great
variation in number of contacts, $R_0$ should be calculated as:
\begin{equation}\label{eq:r0mod}
R_0=c\beta D\left(1+\frac{\sigma^2}{c^2}\right)
\end{equation}
where $\sigma^2$ denotes the variance in the number of
contacts. Another reason that $R_0$ can be an oversimplification is
that most contact networks studied are known to differ significantly
from random interaction.

\section{Network structure and disease dynamics}

 Many contact networks are characterized by a high level of
 transitivity; that is, the number of triangles in the network is much
 larger than is found in an average network having the same frequency
 distribution of number of contacts~\cite{watts}.  A large clustering coefficient
 tends to slow down epidemics because the probability that an infected
 person's contacts will already be infected is very high in such a
 network~\cite{colg,szen}. A common way to estimate clustering in a
 network is to  estimate its relative number of triangles, or more
 exactly, to  calculate the fraction $C$ of all paths of length three
 in the network  which form a triangle:
\begin{equation}\label{eq:c}
C=\frac{3n_\mathrm{triangle}}{n_\mathrm{triple}}
\end{equation}
where $n_\mathrm{triangle}$ is the number of triangles (fully
connected subgraphs of three vertices) and $n_\mathrm{triple}$ is the
number of triples of vertices connected by two or three contacts. The
factor three is needed to normalize $C$ to the interval $[0,1]$.

Another difference between contact and random networks is that most
contact networks are assortative by number of
contacts~\cite{mejn:ass}. This means that individuals who have many
contacts tend to have contact with other individuals who also have
many contacts, and vice versa. The number of contacts is usually
referred to as ``degree'' in network theory, and we will use this term
here. High assortativity decreases the epidemic threshold value among
individuals who have many contacts. The standard measure is the
assortative mixing coefficient $r$; that is, the symmetric correlation
coefficient between the individuals' degrees on each side of all
contacts:
\begin{equation}\label{eq:ass}
r = \frac{4\langle k_1k_2\rangle - \langle k_1+k_2\rangle^2}{2\langle
  k_1^2 +k_2^2\rangle - \langle k_1+k_2\rangle^2}
\end{equation}
where $k_i$ is the degree of the $i$th argument in a list of the
contacts~\cite{mejn:ass}.

\section{Contact networks of patients\label{sec:cont}}

We will now construct networks generated from a unique database
consisting of 295,108 individuals who were registered as
``in-patients'' at any hospital in Stockholm county (pop.\ 1.8
million) during 2001 and 2002.

We want a contact to represent closeness in space and time. Our
spatial requirement is that two patients should be on the same
ward. For the temporal aspect we let closeness be a network parameter,
and we regard a contact between patients as established if they were
hospitalized on the same ward for a duration $t_\mathrm{ol}$ (overlap
time) or longer. A $t_\mathrm{ol}=0$ means that contacts between one
patient who was admitted the same day as another patient was
discharged are included. Furthermore, we let the sampling time window
size $\Delta t$ be another parameter. The two parameters
$t_\mathrm{ol}$ and $\Delta t$ yield different networks that are
relevant to different diseases. For example, for diseases such as
measles, SARS, and norovirus, which spread rapidly, a narrow time
window will be appropriate~\cite{hey}; for diseases requiring
prolonged contact for transmission, like tuberculosis, the relevant
network is represented by a larger $t_\mathrm{ol}$.

\begin{figure}
  \resizebox*{0.85\linewidth}{!}{\includegraphics{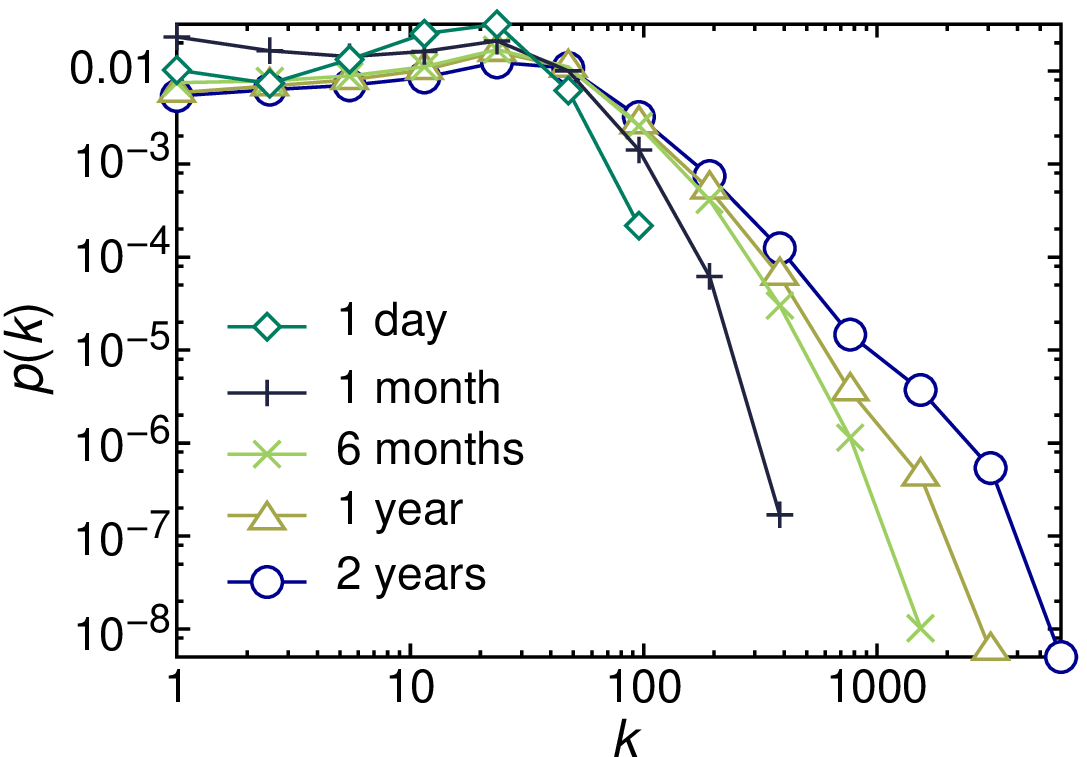}}\\
  \resizebox*{0.85\linewidth}{!}{\includegraphics{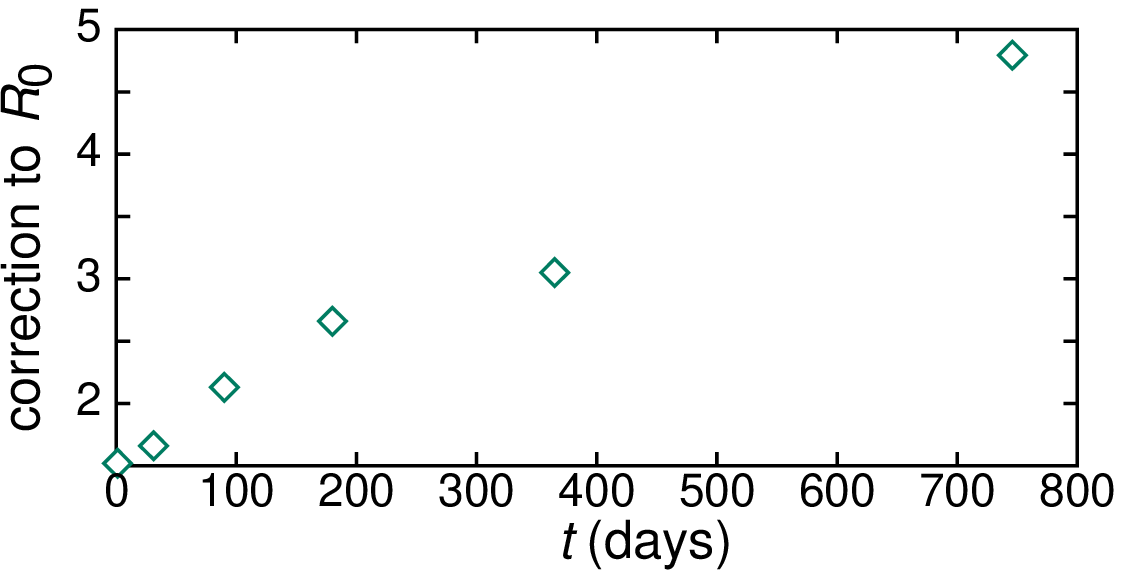}}
  \caption{ The degree distribution and its effect on the reproduction
    number. In the upper panel we see the probability density function
    $P(k)$ versus $k$ (with logarithmic binning) for networks with
    overlap $t_\mathrm{ol} = 0$ and different time windows. The lower
    panel shows the correction to the basic
    reproduction number $R_0$ as a function of the time window size
    for $t_\mathrm{ol} = 0$. For example, for $\Delta t = 2$ years an
    epidemic can occur by a disease five times less transmissible than
    predicted by traditional models.
  }
  \label{fig:deg}
\end{figure}

\section{Network structural properties}

\subsection{Degree distribution}

We will start by plotting the probability density function $P(k)$ for
an individual to have $k$ contacts. This function is plotted for,
respectively:
\begin{itemize}
\item One weekday in January.
\item The entire month of January.
\item The first six months of 2001.
\item The whole period 2001-2002.
\end{itemize}

\begin{figure}
  \resizebox*{0.9\linewidth}{!}{\includegraphics{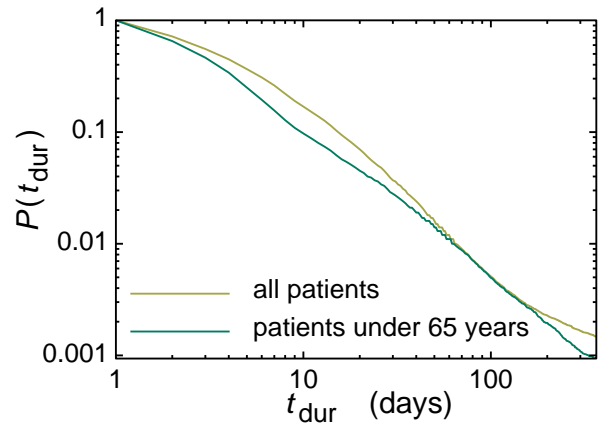}}
  \caption{ The probability that any hospitalization lasts
    $\Delta t$ or longer. In A, all data are used; in B, the
    patients over 60 years are removed from the data set. 
  }
  \label{fig:hosp}
\end{figure}

In Figure~\ref{fig:deg}A we see the development of this contact
structure from an exponential distribution to a degree distribution
with a truncated ``fat tail.'' It is clear that variation in the
number of contacts between individuals increases with
time. Figure~\ref{fig:deg}B shows the degree to which calculated using
Equation~\ref{eq:r0} must be compensated according to
Equation~\ref{eq:r0mod} for this increase in variation. The skewed
degree distribution in our case is related to the power-law-like
distribution of hospitalization times (see
Figure~\ref{fig:hosp} and Sect.~\ref{sect:hosp}). The distribution of hospitalization will
indirectly lead to ``preferential attachment''~\cite{ba:model,klemm}
(that is, a heightened probability of high-degree vertices to form
additional contacts)---a well known mechanism for producing fat-tailed
degree distributions~\cite{price}.

\begin{figure*}
  \resizebox*{0.4\linewidth}{!}{\includegraphics{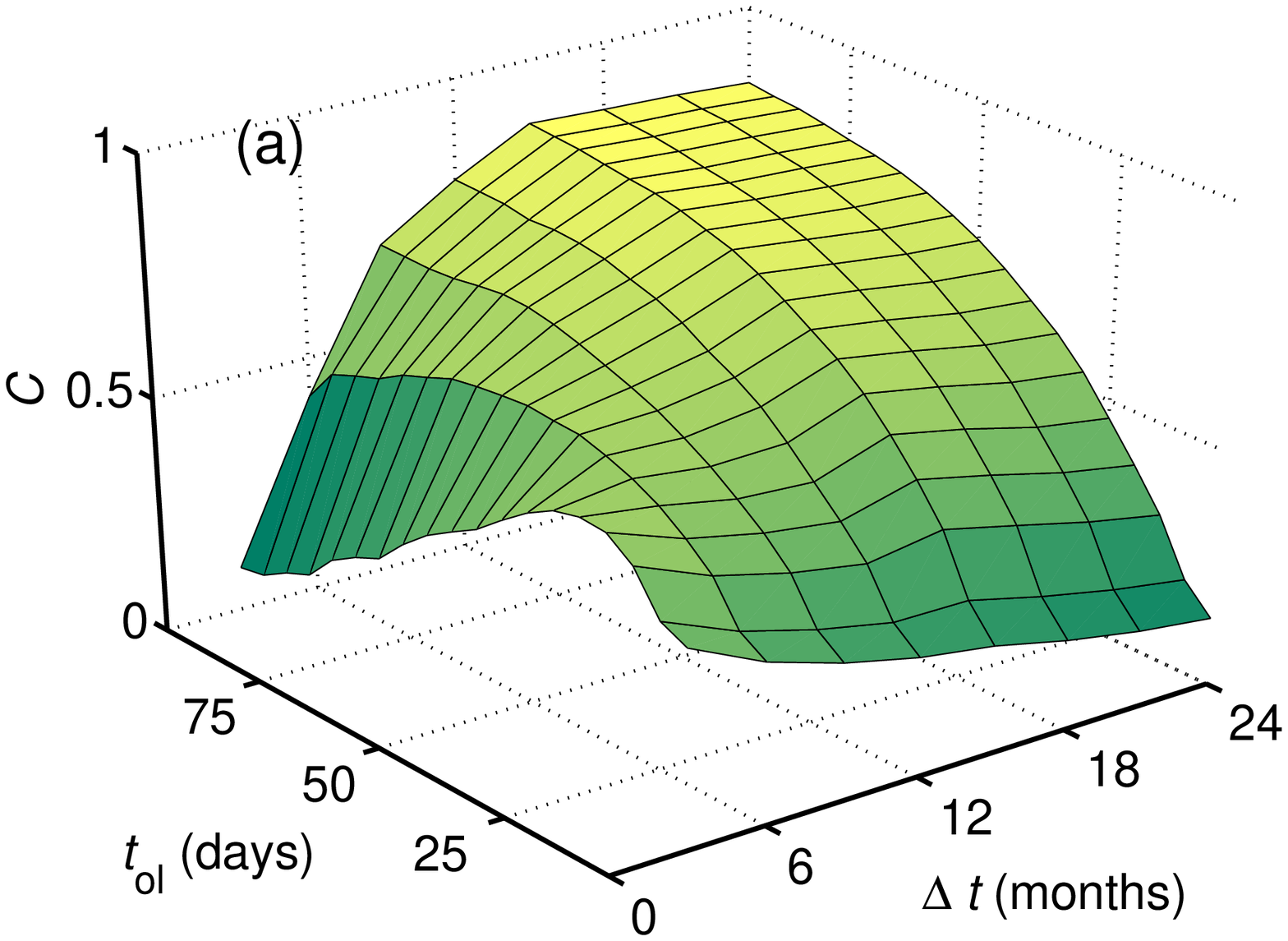}}~
  \resizebox*{0.4\linewidth}{!}{\includegraphics{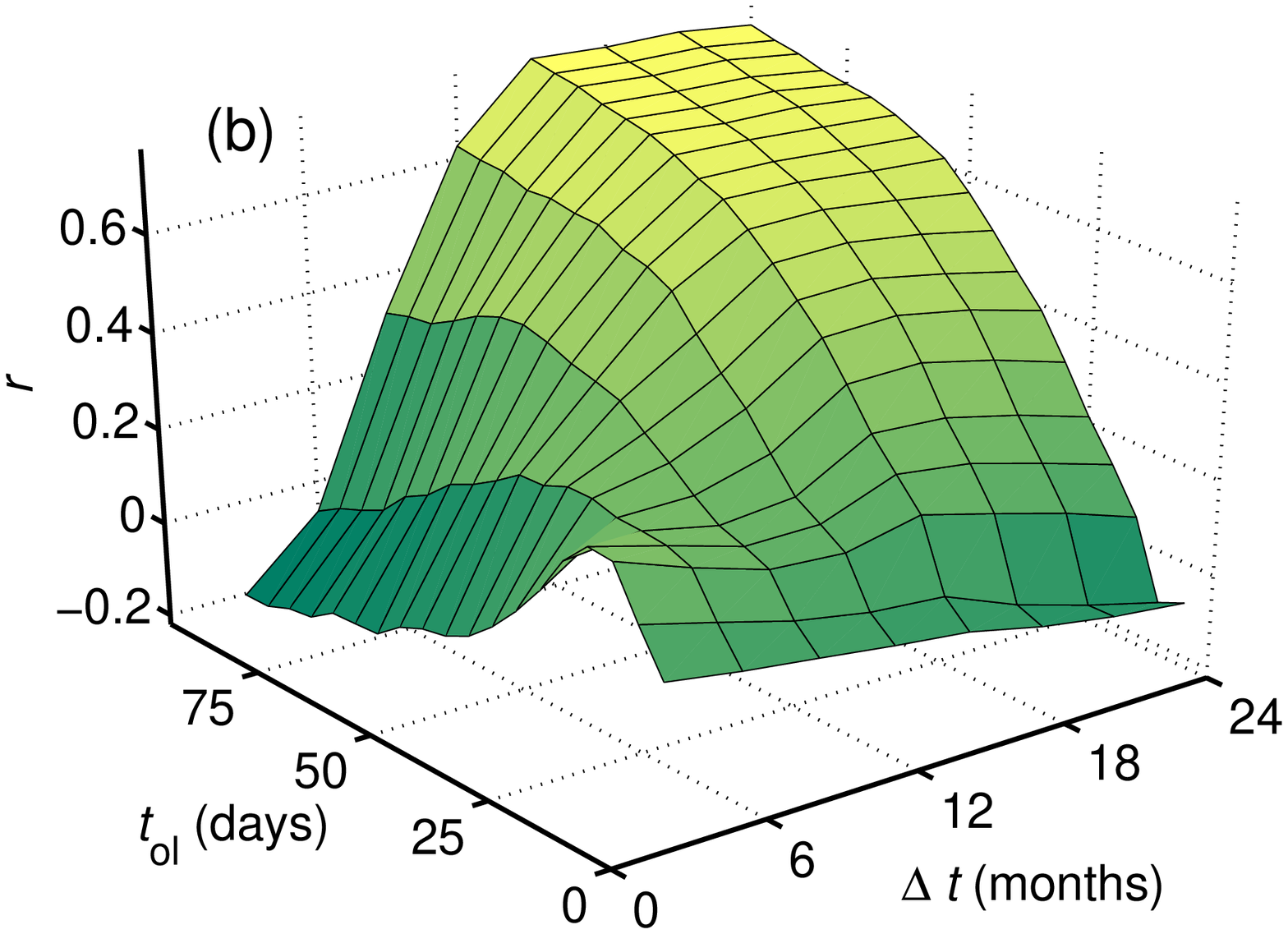}}
  \caption{
    The clustering (A) and assortative mixing (B) coefficients, $C$
    and $r$, as functions of sampling time, $\Delta t$, and overlap
    time, $t_\mathrm{ol}$.
  }
  \label{fig:cr}
\end{figure*}

\subsection{Transitivity and assortativity}

We will now investigate how transitivity, $C$, and assortativity, $r$,
vary with sampling time, $\Delta t$, and overlap time,
$t_\mathrm{ol}$.

In Figure~\ref{fig:cr}(A) we display the clustering coefficient, $C$,
and in Figure~\ref{fig:cr}(B) the level of assortative mixing, $r$, as
functions of our time intervals $t_\mathrm{ol}$ and $\Delta t$. We
note that both parameters exhibit a very similar functional form. $C$
and $r$ both decrease with $\Delta t$ when is held constant for small
$t_\mathrm{ol}$, butincrease when tol is held constant for large
values of $t_\mathrm{ol}$. Both parameters behave in a similar way
when $\Delta t$ is held constant. For small values of $t_\mathrm{ol}$,
the parameters first increase and then start to decrease as a function
of $t_\mathrm{ol}$. For large values of $t_\mathrm{ol}$, both
parameters first increase with $t_\mathrm{ol}$ until they converge at
a high level of clustering and assortativity.

The estimated high levels for the $C$ and the $r$ parameters, and the
resemblance in functional form between the $r$-surface and the
$C$-surface in Figure~\ref{fig:cr} are a consequence of the
compartmental structure of the healthcare system. Assume a
hypothetical network, $\Omega$ in which all inpatients stay on the
same ward during the entire duration of
$\Delta t$. In such a network, each inpatient will have a link to each
other inpatient on the same ward. The level of clustering between the
inpatients will therefore equal 1, both locally on each ward and
globally throughout the whole healthcare system, because no links
exist between inpatients on different wards in $\Omega$. It is trivial
that the level of assortativity can only be defined if there is a
variation in ward size, and that $r$ in these cases must equal 1
because the degree of contacts on each side of each link will be
equal. Our results show that networks defined by a large value of
$t_\mathrm{ol}$ and a large value of $\Delta t$ come very close to
$\Omega$.  Both $C$ and $r$ are large, and the vast majority of
individuals in these networks are registered as inpatients only once
per ward (see supplementary material). If we relax the restrictions on
$\Omega$ such that each inpatient stays on the same ward during the
entire period of $\Delta t$, the $C$ and $r$ parameters may drop below
1. This makes it possible for triples, which not are triangles, to be
formed between inpatients that stay on different wards, and between
inpatients that stay on the same wards but at different times. This
also makes it possible for links to form between nodes with different
degrees. The same occurs where $\Delta t$ increases when
$t_\mathrm{ol}$  is held constant for low levels of
$t_\mathrm{ol}$. The decay in $C$ and $r$ is a result of the skew
distribution of hospitalization times (Fig.~\ref{fig:hosp})---a
long-term patient A will form a triple (but not a triangle) with the
many pairs of short-term patients whose hospitalization does not
overlap with each other's but does overlap with A. This situation, and
the number of inpatients who stay on more than one ward, will be more
common for larger time windows, causing $C$, and consequently $r$, to
decay over time.

That C increases with $t_\mathrm{ol}$ (for fixed $\Delta t$) may seem
counterintuitive: As $t_\mathrm{ol}$ grows, the network will have
fewer edges, and also fewer triangles. We have constructed a simple
agent-based model that shows that a prerequisite for this is the
observed skewed distribution of hospital stays (see
Sect.~\ref{sec:model}).

\section{Summary and conclusions}

Our study of a very large inpatient database shows that hospital
systems characteristically have a very large variation in duration of
hospital stays, which generates a correspondingly large variation in
number of contacts. We have further shown that the clustering
coefficient and assortative mixing depend greatly on sampling time and
the length of time that two inpatients must spend together for contact
to be effected. Both of these coefficients, $C$ and $r$, become
extremely large in our real-life network when $t_\mathrm{ol}$ and
$\Delta t$ are large. This is alarming because it has been shown that
both a high level of clustering and a high level of assortative
interaction decrease the epidemic
threshold~\cite{mejn:clu,mejn:ass}. Any strategy to intervene with
disease spread in a hospital environment must take into account this
departure from assumptions of random interaction and homogenous
mixing. For infections with high transmissibility, short incubation
times and short duration of infectiousness, such as norovirus
infections and SARS, our finding may be less
important. However, for diseases such as tuberculosis or MRSA
characterized by low transmissibility and long duration of
infectiousness, it becomes necessary to take this variation into
consideration because a high variance will lower the epidemic
threshold.

Our findings indicate that the individuals with many contacts are
significant for the spread of infectious diseases with long duration
of infectiousness. These high-risk individuals will probably be
identifiable through hospital patient registration systems, and should
be the first to be targeted by contact tracing. The high level of
clustering further indicates that it may be worth screening all
inpatients that have spent time on the same ward as positive
inpatients before and after the positive inpatients were there. The
high level of clustering makes it reasonable to assume that more than
one inpatient will be infectious at the same time on the same ward,
and consequently that the disease would have existed among the
inpatients on the ward both before and after the actual inpatient in
question was on the ward.

\begin{acknowledgments}
This research is supported by The Swedish Emergency Management Agency,
European research NEST project DYSONET 012911. The data has generously
been made available by Stockholm County Council with help from Torsten
Siegel. The study was conducted with permission from the Regional
Ethical Review Board in Stockholm (record 2004/5:8).
\end{acknowledgments}

\appendix

\begin{figure}
  \resizebox*{0.8\linewidth}{!}{\includegraphics{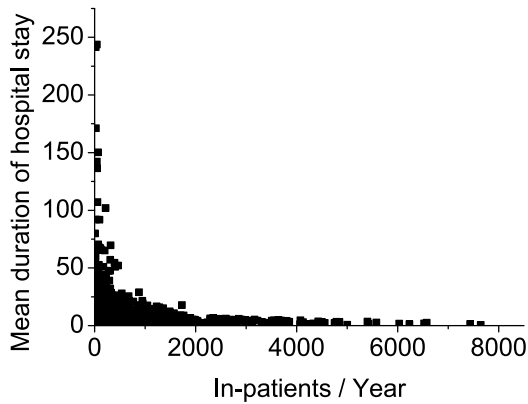}}
  \caption{Mean duration of hospital stay for the inpatients for each
    ward versus number of inpatients a year for the wards in the
    database.
  }
  \label{fig:dur}
\end{figure}

\begin{figure*}
  \resizebox*{0.4\linewidth}{!}{\includegraphics{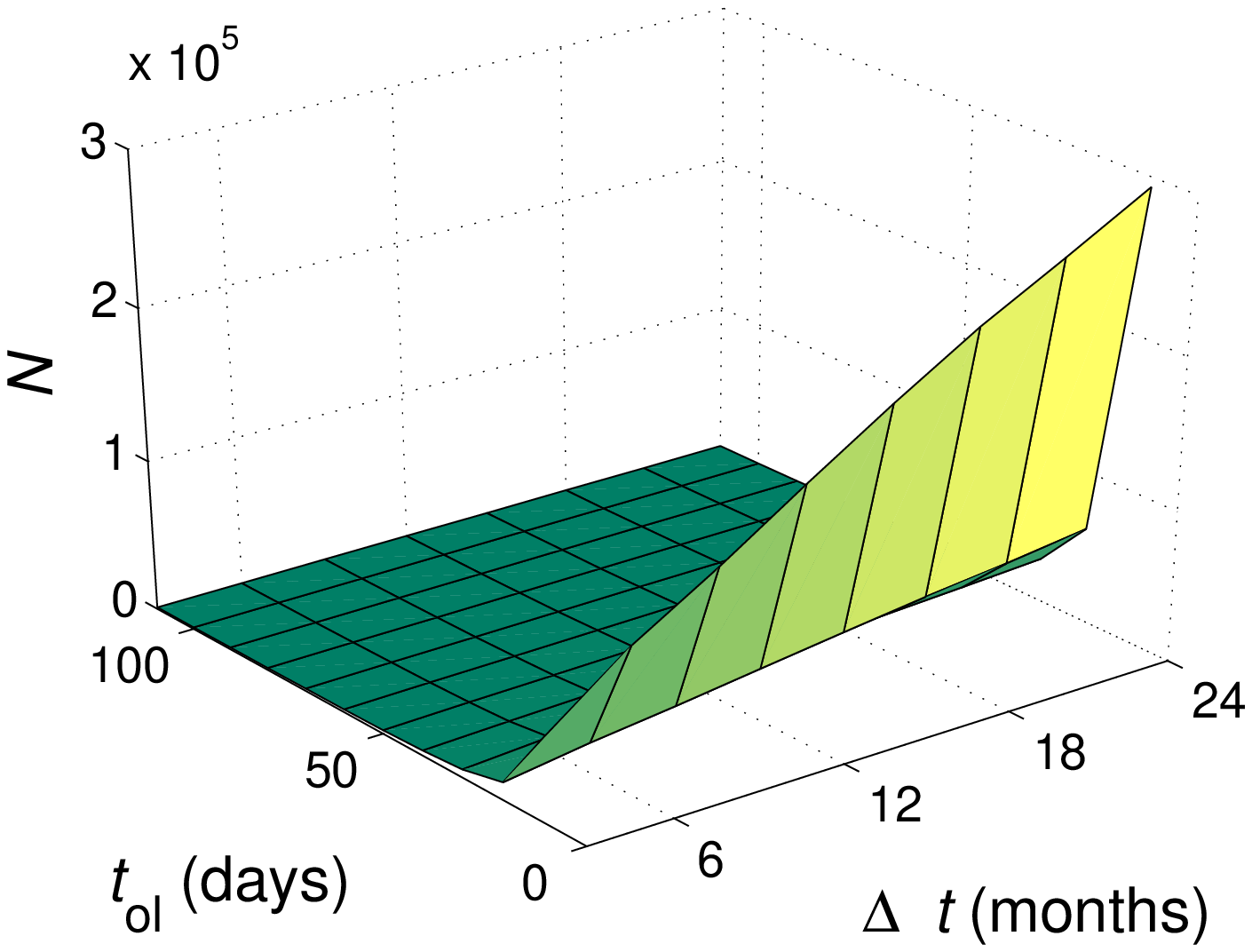}}~
  \resizebox*{0.4\linewidth}{!}{\includegraphics{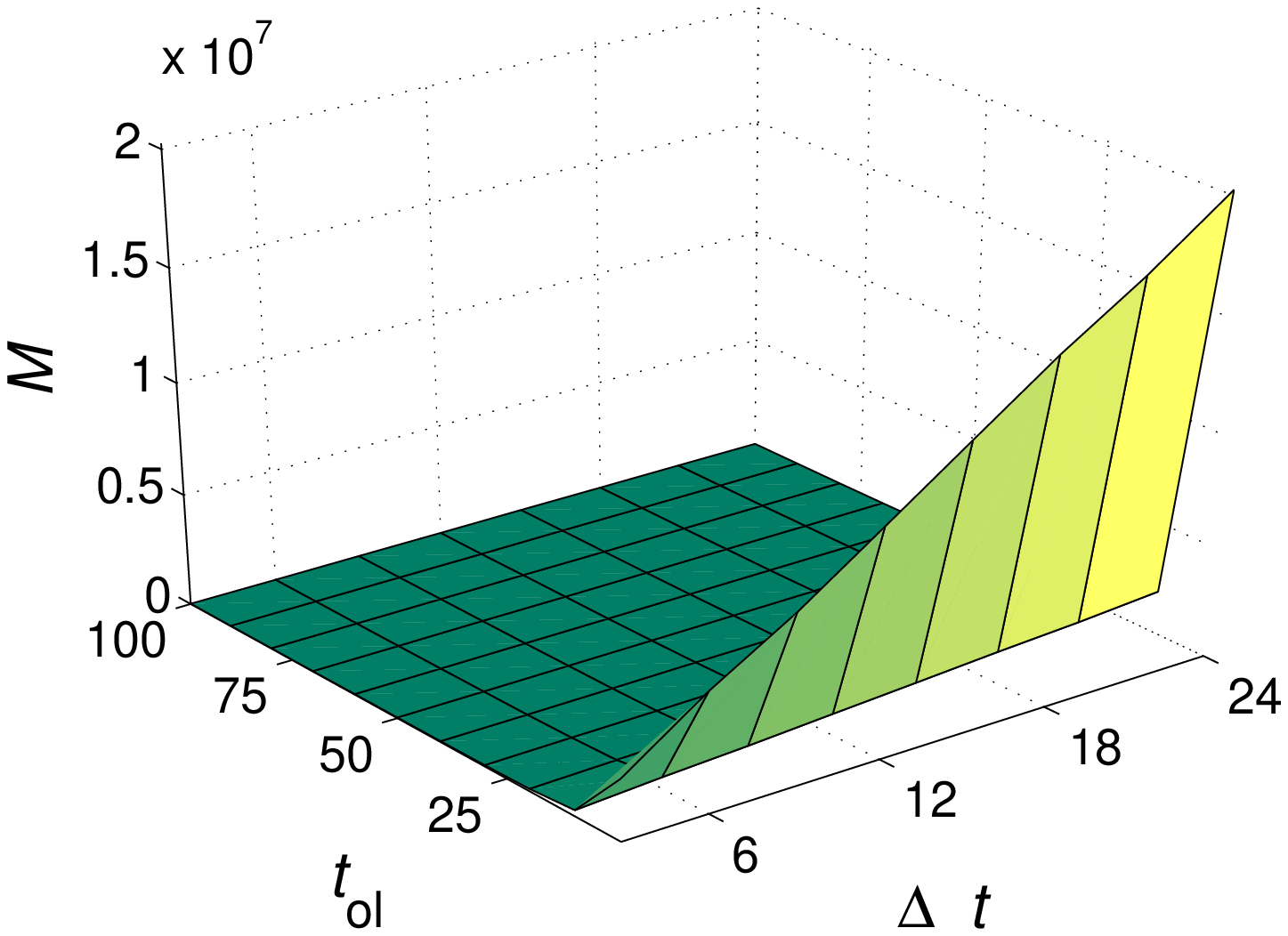}}
  \caption{The number of vertices (A) and number of edges (B), $N$ and
    $M$, as functions of the overlap time $t_\mathrm{ol}$ and time
    window size $\Delta t$.
  }
  \label{fig:nm}
\end{figure*}

\begin{figure}
  \resizebox*{0.9\linewidth}{!}{\includegraphics{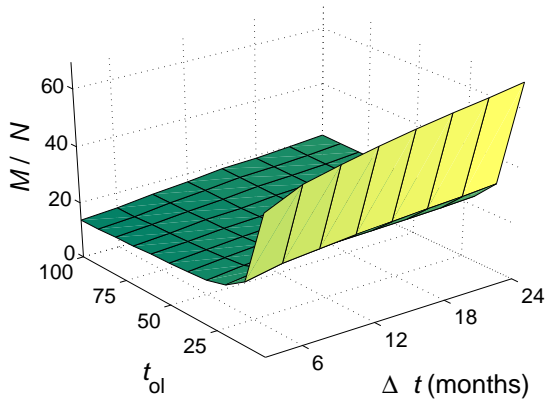}}
  \caption{The average number of edges per vertex,  $N/M$, as
    functions of the overlap time $t_\mathrm{ol}$ and time window size
    $\Delta t$.
  }
  \label{fig:n_m}
\end{figure}

\begin{figure}
  \resizebox*{0.45\linewidth}{!}{\includegraphics{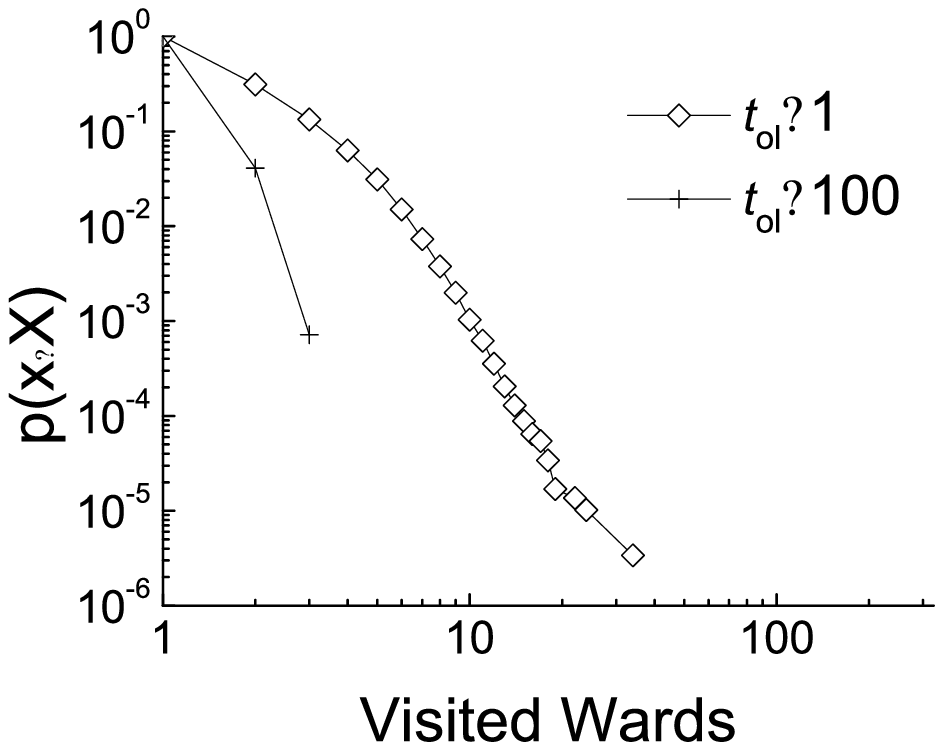}}
  \resizebox*{0.45\linewidth}{!}{\includegraphics{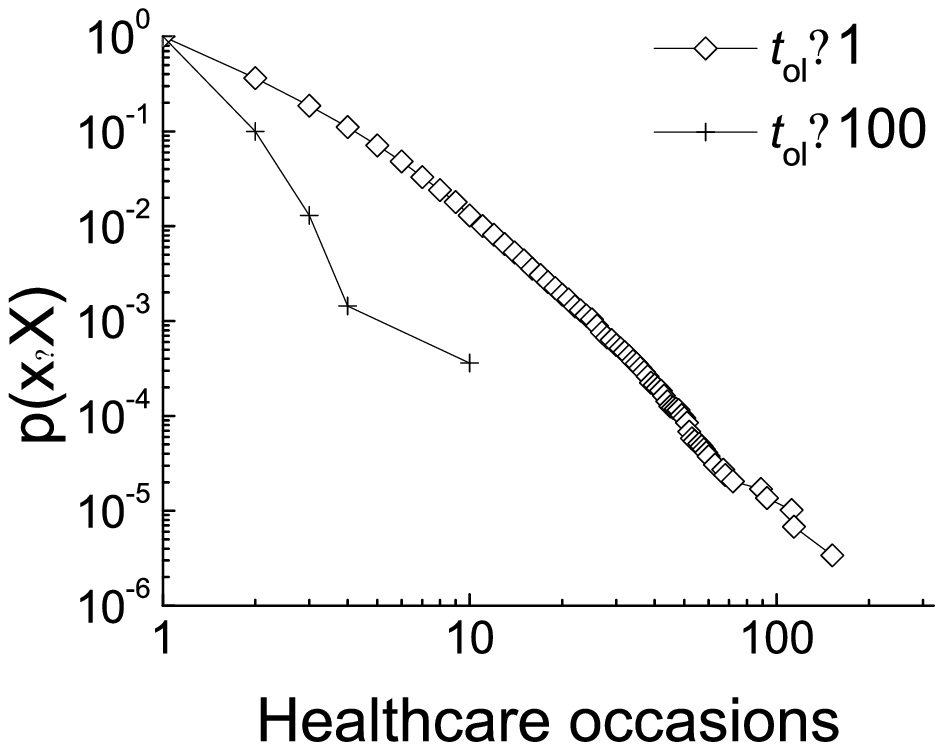}}
  \caption{The cumulative distribution, $p(x\geq X)$, for the number
    of healthcare occasions per inpatient and the number of visited
    wards per inpatient during the period 2001-2002.
  }
  \label{fig:hc_occ}
\end{figure}

\section{Further statistics}

The dataset contains information for 570,382 ward admissions,
including date of admission, date of discharge, and ward
identity. There were a total of 702 wards located at 52 different
geographical units such as hospitals and nursing homes. The mean daily
number of patients on the wards for the two-year period varied between
1 and 69 (mean 10.05, SD 9.44). Wards with a large number of
inpatients per year strongly tend toward shorter duration of inpatient
hospital stays, and vice versa, as shown in Figure~\ref{fig:dur}.

As described in the Sect.~\ref{sec:cont}, we define a network as the
individuals who have been inpatients some time during the sample time,
$\Delta t$, and a contact as a link between two individuals who have
been inpatients on the same ward for a duration $\geq
t_\mathrm{ol}$. Figure~\ref{fig:nm}A and \ref{fig:nm}B show the number
of nodes, $N$, and the number of links, $M$, for nodes having at least
one link with a duration $\geq t_\mathrm{ol}$.

The $N$ and $M$ surfaces show a large variation in absolute size. The
surface for the number of vertices per node shows a similar
surface. The quote between the largest value and smallest value is,
however, smaller by several orders of magnitude.

The number of healthcare occasions and number of different wards
visited during the period under study varies a widely for different
values of, $t_\mathrm{ol}$ (see Figure~\ref{fig:hc_occ}). The number
of separate healthcare occasions for all contacts, that is, to
$t_\mathrm{ol}\geq 1$, in particular exhibits a fat tail. This holds
to some extent for the number of visited wards as well. The
individuals who had at least one contact with a duration of at least
100 days are thus considerably less mobile between the wards in the
hospital system than those who have not.

The dataset is associated with one known systematic bias in the sense
that one single inpatient may be registered as an inpatient on two
wards at the same time such as when an inpatient is moved for a short
period but is expected to return. Our analyses show that one single
individual is registered on two separate wards 6734 times. We have not
been able to show that these biases have any significant effect on the
results we are presenting in this paper and will therefore use the
whole dataset in our analyzes.

\begin{figure}
  \resizebox*{0.9\linewidth}{!}{\includegraphics{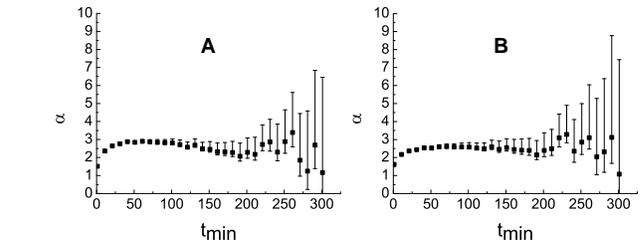}}
  \caption{The best estimates of the slope  for different values of
    for the whole population (A) and for individuals younger than 65
    years old (B) The error bars are 95\% confidence intervals
    generated by bootstrapping.
  }
  \label{fig:expo}
\end{figure}

\section{Notes on the distribution of hospitalization
  times\label{sect:hosp}}

In Figure~\ref{fig:expo} we have plotted the cumulative distribution,
of $t_\mathrm{dur}$, for all healthcare occasions in 2001. This allows
us to plot the cumulative distribution in the interval 1 to 365 days
for all of these healthcare occasions without interference from any
finite size effects of the material in this interval. The plot shows
that the duration of hospital stays exhibits a skewed power-law-like
tail, $p(t_\mathrm{dur})\sim t_\mathrm{dur}^{-\alpha}$. We estimate
the slope, $\alpha$, in the interval $[t_\mathrm{min}, 365]$ by
fitting $\alpha$ in $p(t_\mathrm{dur})=t_\mathrm{dur}^{-\alpha}/\tau$,
where
\begin{equation}\label{eq:tau}
\tau = \sum_{i=t_\mathrm{min}}^{365} i^{-\alpha}
\end{equation}
is a normalization factor. A maximum likelihood procedure was used for
the estimation.  The 95\% confidence intervals were estimated by
bootstrapping. Figure~\ref{fig:expo}A and \ref{fig:expo}B show how
$\alpha$ changes when $t_\mathrm{min}$ is increased.

\section{A model of contact networks of patients\label{sec:model}}

\begin{figure}
  \resizebox*{0.95\linewidth}{!}{\includegraphics{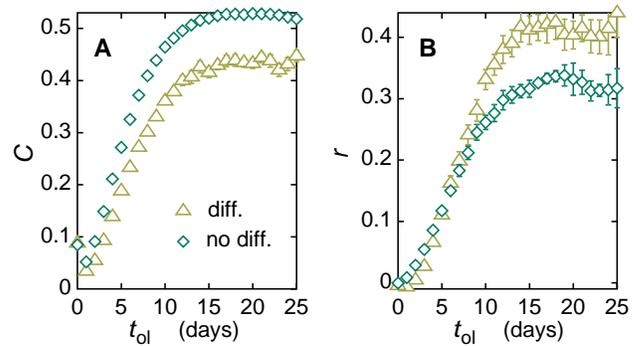}}
  \caption{he assortative mixing (A) and clustering coefficients (B)
    as functions of the overlap time $t_\mathrm{ol}$ for a simulated
    healthcare system. For the ``no diff.'' curves, patients are
    assigned to a random ward, whereas for "diff." curves, patients
    with a similar duration of hospital stay share wards (which
    reproduce the functional forms of Fig. 3). The simulation
    parameters are $N = 10000$, $N_w = 50$, $p_1 = 0.02$ / day, $p_2 =
    1 / 3$, $\Delta t = 2500$ days, and $P_t\sim t^{-3}$. The curves
    are averaged over 10 runs of the algorithm.
  }
  \label{fig:model}
\end{figure}

To answer the question why $C$ increases with $t_\mathrm{ol}$ (for
fixed $\Delta t$) we construct a simple agent-based model of a
healthcare system from first principles: Suppose a healthcare system
of Nw wards of equal capacity is intended to serve a population of $N$
individuals. Each day a non-hospitalized individual  hospitalized with
a probability $p_1$ and will stay for a random time $t$ (of some
distribution $P_t$) on ward $w$ (how the ward is chosen is discussed
below). After hospitalization the patient is either transferred to
another ward with probability $p_2$ for a duration of a new t or
discharged. This dynamic, given a $\Delta t$ and $t_\mathrm{ol}$,
yields networks just like our real data did.

How shall we assign patients to the wards? The simplest assumption is
to choose the wards with equal probability. As seen in
Figure~\ref{fig:model}, this yields the shape of $C$ and $r$ seen in
Figure~\ref{fig:cr}.

One important feature is missing from this model: different specialty
wards hospitalize patients for different durations. If we incorporate
this, the curves stay qualitatively the same. From the model, we
understand that for large overlap times the long-term patients form
densely connected components---otherwise the network is empty. The
model is insensitive to parameter values. A skewed $P_t$ function is,
however, needed. 

The algorithm consists of the following steps repeated t times (that
is, one of these steps corresponds to one day in the simulation):
\begin{enumerate}
\item Go through all healthy (non-hospitalized) patients. With a
  probability $p_1$ hospitalize a patient. The duration of the
  hospitalization is given by $P_t$. Assign a ward according to the
  methods listed below. In our simulation, we choose $P_t \sim
  t^{-3}$. This is based on the observed distribution of
  hospitalization times (see Figure~\ref{fig:hosp}).
\item Go through all newly discharged patients. With probability $p_2$
  re-hospitalize a patient. The duration of the hospitalization is
  given by $P_t$. Assign a ward according to the methods listed below.
\item If needed, construct the network according to the method
  detailed in the Sect.~\ref{sec:cont}.
\end{enumerate}

To assign a ward given a hospitalization time, we propose two
different methods. The first option is to select the ward by uniform
randomness. Clearly, this will, on average, make all wards equally
full. This method is used for the ``no diff.'' curves in
Figure~\ref{fig:hosp}. However, the hospitalization times are very
different---on some wards, the hospitalization time is much longer
than average; on others, people stay for very short periods. To model
this, we differentiate strictly between the wards so that the patients
on ward $w_i$ always stay a shorter time than the patients on ward
$w_{i+1}$. We implement this by generating $N_\mathrm{rnd}$ random
numbers distributed according to $P_t$. Then we sort these values in
increasing values of $t$ and associate ward 1 with the $t$-values
$[t_1,\ldots,t_{s1}]$, ward 2 with the $t$-values
$[t_{s1}+1,\ldots,t_{s2}]$, and so on, such that the sum of $t$-values
are the same for all wards. During the iterations, a random value of
the array of random numbers is drawn and the patient is assigned to
the corresponding pre-assigned ward. We use $N_\mathrm{rnd} = 10^6$
for Figure~\ref{fig:model}. The same plot with $N_\mathrm{rnd} = 10^4$
yields indistinguishable curves. This is the method used for the
``diff.'' curves in Figure~\ref{fig:model}. Finally, to obtain the
curves in  Figure~\ref{fig:model}, we average the result of 20 runs of
the algorithm above.

\end{document}